\documentclass{PoS}

\title{Reconstruction of physics objects in the CMS detector}

\ShortTitle{Reconstruction of physics objects in the CMS detector}

\author{\speaker{Aruna Kumar Nayak}\thanks{On behalf of CMS Collaboration}\\
        IRFU, CEA, Saclay, France\
        E-mail: \email{Aruna.Nayak@cern.ch}}

\abstract{The reconstruction and identification of physics objects in the CMS detector, in the context of the charged Higgs boson search analysis, are presented. The reconstruction algorithms and their performance in 7 TeV and 8 TeV LHC data are discussed. The identification of tau hadronic decays,  the reconstruction of hadronic jets and missing transverse energy and the identification of b jets are described in detail.
}

\FullConference{Prospects for Charged Higgs Discovery at Colliders - Charged 2012,\\
		October 8-11, 2012\\
		Uppsala University, Sweden}

\begin{document}

%\section{...}
\section{Introduction} 
Some extensions of the Standard Model of Particle Physics, such as the Minimal Supersymmetric Standard Model (MSSM) or the Two Higgs Doublet Model, predict the presence of charged Higgs bosons. 
For large values of tan$\beta$, the ratio of the vacuum expectation values of the two Higgs doublets, the charged Higgs decays to a tau lepton and a neutrino. 
The CMS experiment at CERN \cite{JINST:3:S08004} has performed the search for a light charged Higgs boson decaying to $\tau\nu_{\tau}$ using 7 TeV LHC data \cite{HIG_11_019}. 
The search requires efficient identification of $\tau$-leptons as well as a good reconstruction of hadronic jets, missing transverse energy and a good b jet identification.  

The CMS experiment has developed excellent techniques for the reconstruction and identification of physics objects. 
In this article, the CMS reconstruction and identification techniques are briefly discussed in order of their importance to the charged Higgs search analysis. 
Many physics analyses at CMS use high level physics objects reconstructed using the particle-flow event reconstruction technique. A brief description of the particle-flow event reconstruction is presented below.  

\section{Particle-Flow Event Reconstruction}
The particle-flow event reconstruction algorithm aims at reconstructing all stable particles in the event by combining information from all CMS sub-detectors. 
The algorithm optimizes the determination of particle types, directions and their energies.
The resulting list of particles are then used to reconstruct higher level objects such as jets, taus, missing transverse energy, to compute charged lepton and photon isolation, etc.
The details of the particle-flow event reconstruction developed at CMS can be found in \cite{PAS-PFT-09-001}. 

The basic elements of the particle-flow event reconstruction are the charged particle tracks reconstructed in the central tracker and the energy clusters reconstructed in electromagnetic and hadronic calorimeters.
The charged particle tracks are reconstructed using an iterative tracking strategy, described in \cite{PAS-PFT-10-002}, with both a high efficiency and a low fake rate for charged particle momentum as low as 150 MeV/$c$. 
The energy clustering is performed in each sub-detector of the calorimeters separately using a specific clustering algorithm, developed for particle-flow event reconstruction, which aims for a high detection efficiency even for low energy particles and separation of close energy deposits.   
These basic elements are then connected to each other using a link algorithm to fully reconstruct each single particle, while removing any possible double counting from different detectors.  
The algorithm produces $``$blocks$"$ of elements linked directly or indirectly. 
The particle-flow algorithm described in \cite{PAS-PFT-10-002} is finally used to reconstruct and identify a set of particles from each block of elements. 
Charged hadrons are reconstructed from the tracks in the central tracker.
Photons and neutral hadrons are reconstructed from energy clusters in calorimeters. Clusters separated from the extrapolated position of tracks in the calorimeters constitute a clear signature of these neutral particles.
A neutral particle overlapping with charged particles in the calorimeters can be detected as a calorimeter energy excess with respect to the sum of the associated track momenta. 
The resulting list of reconstructed particles constitute a global description of each event, available for subsequent physics analysis. 

\section{Hadronic Jets} %1 page
Quarks and gluons produced in hard scattering of partons in pp collisions manifest themselves as hadronic jets. 
A detailed understanding of the jet energy calibration and resolution is of crucial importance and is a leading source of uncertainty for many analyses with jets in the final state.  
A brief description of CMS jet reconstruction algorithms, jet energy calibration techniques and the jet energy scale uncertainties are presented here. 
  
\subsection{Reconstruction of Hadronic Jets}
Four types of jets are reconstructed at CMS depending on the input to the jet clustering algorithm: calorimeter jets, Jet-Plus-Track (JPT) jets, Particle-Flow (PF) jets, and track jets~\cite{PAS-JME-10-003}. 
Jets presented here are reconstructed using the anti-$k_{T}$ \cite{antiKt} clustering algorithm with the size parameter $R ~=~ 0.5$. 
To evaluate their performance, in Monte Carlo simulations, generated jets (GenJets) or particle jets are reconstructed as well by applying the same jet clustering algorithm to all stable generated particles. 
Since most of the analysis at CMS are using PF jets, we concentrate here on PF jets while briefly describing two other types of jets.

{\it Calorimeter jets} are reconstructed using energy deposits in the calorimeter towers, where calorimeter tower consists of one or more hadronic calorimeter (HCAL) cells and the geometrically corresponding electromagnetic (ECAL) crystals. 
The {\it Jet-Plus-Track} algorithm \cite{PAS-JME-09-002} exploits the excellent performance of the CMS tracking detectors to improve the $p_{T}$ response and resolution of calorimeter jets.   
For each track in the jet, the average expected calorimeter energy is subtracted and the momentum measured in the tracker is added to the jet. For the tracks which are bent out of the jet cone due to magnetic field the momentum of the track is added to the jet. 
PF jets are reconstructed from the list of particles reconstructed using particle-flow algorithm. 
The jet momentum and spacial resolutions are improved with respect to the calorimeter jets, since the use of tracking detectors and excellent ECAL granularity allows to resolve and precisely measure charged hadrons and photons inside jets.   

\subsection{Energy Calibration of Hadronic Jets}
\label{ChapterJEC}
Due to the non-uniform and non-linear response of the CMS calorimeters the jet energy measured in the detector is typically different from the corresponding particle jet energy. 
Furthermore, electronic noise and additional pp interactions in the same bunch crossing (event pile-up) leads to extra unwanted energy.
The purpose of the jet energy correction is to relate, on average, the energy measured for the detector jet to the energy of the corresponding particle jet. 

\begin{figure}[htp]
%\begin{center}
\centering
\begin{tabular}{c}
\includegraphics[width=0.45\textwidth]{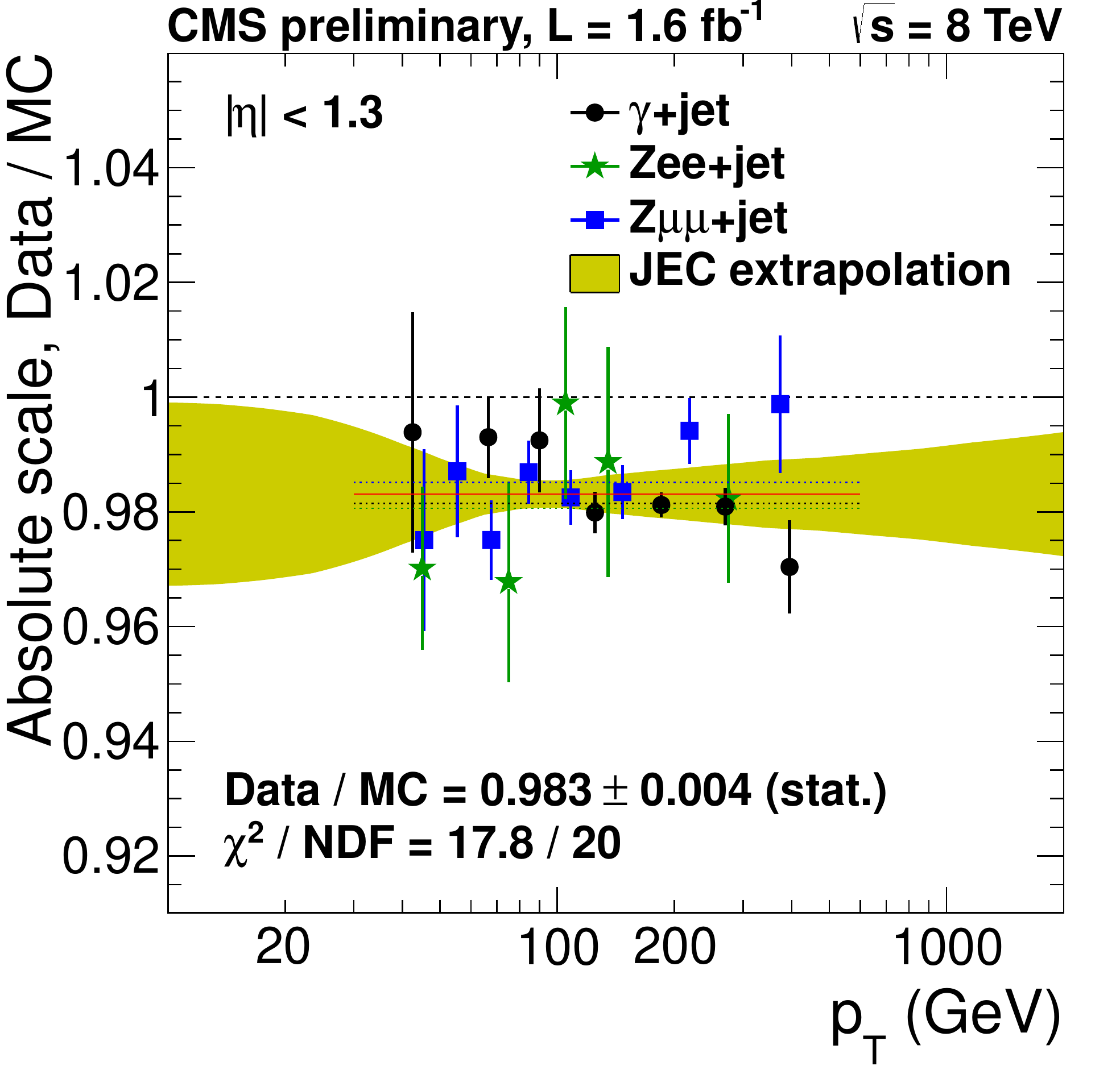} \hfill
\includegraphics[width=0.45\textwidth]{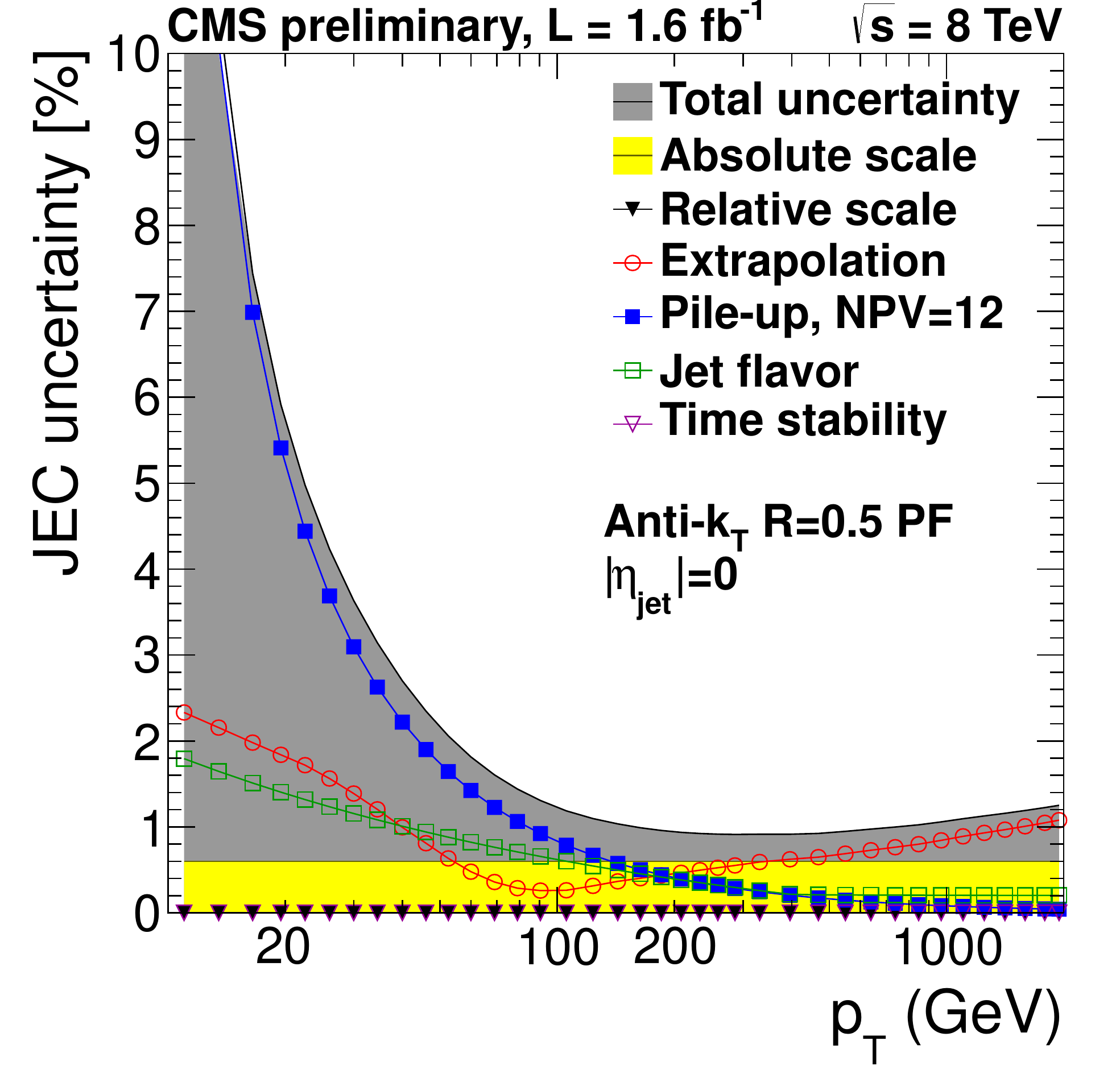} 
\end{tabular}
\caption{Left: Data/MC ratio of absolute jet energy scale as function $p_{T}$ of reference object.
Right: Difference sources of JEC uncertainties as function of jet $p_{T}$. 
}
\label{fig:JEC}
%\end{center}
\end{figure}

CMS has developed a factorized multi-step procedure for the jet energy calibration (JEC) \cite{JME-10-011}. The correction is applied as a multiplicative factor to each component of the raw jet four momentum vector $p_{\mu}^{raw}$ as shown in Eq.\ref{eqn:JEC}.
\begin{eqnarray}
p_{\mu}^{corrected}~=~p_{\mu}^{raw}~\cdot~C_{offset}(p_{T}^{raw})~\cdot~C_{MC}(p_{T}^{'}, \eta)~\cdot~C_{rel}(\eta)~\cdot~C_{abs}(p_{T}^{''})
\label{eqn:JEC}
\end{eqnarray}  
   
where $p_{T}^{'}$ is the transverse momentum of the jet after applying offset correction and $p_{T}^{''}$ is the transverse momentum of the jet after all previous corrections. 
$C_{offset}$ is the offset correction derived using the jet area method \cite{JME-10-011}. For each event, an average $p_{T}$ density $\rho$ per unit area is estimated, which characterizes the soft jet activity and is contamination of the underlying event, the electronic noise and the pile-up. 
The MC calibration, $C_{MC}$, is based on the simulation and corrects the energy of the reconstructed jets such that it is equal to the energy of generated MC particle jets. 
It removes the bulk of the non-uniformity in $\eta$ and the non-linearity in $p_{T}$.  
The residual corrections $C_{rel}$ and $C_{abs}$ for the relative and absolute energy scales, respectively, are derived using data driven method, using dijet and $\gamma/Z$+jets events, to account for the small differences between data and simulation. 
Fig.~\ref{fig:JEC} (left) shows the data/MC ratio for the absolute jet energy scale as function of $p_{T}$ of the reference object. 
The uncertainty due to various sources are shown in Fig.~\ref{fig:JEC} (right) as function of jet $p_{T}$. The uncertainty due to pile-up dominates at low $p_{T}$.     

\section{Reconstruction of Hadronic Tau Decays} %3 page
The $\tau$ lepton is the heaviest among the three Standard Model leptons and has a very short lifetime. Thus it decays before reaching the detector elements. In two thirds of the cases, $\tau$ leptons decay hadronically, typically into one or three charged mesons (predominantly $\pi^{+}$, $\pi^{-}$), often accompanied by neutral pions and a $\nu_{\tau}$. The algorithm designed at CMS uses particle-flow particles, such as final state photons and charged hadrons, to identify hadronic decays of $\tau$ leptons ($\tau_{h}$) through the reconstruction of the intermediate resonances. The $\nu_{\tau}$ escapes undetected and is not considered in the $\tau_{h}$ reconstruction. 

\subsection{$\tau_{h}$ Identification Algorithm}
The algorithm used at CMS to reconstruct and identify hadronic $\tau$ decays is called Hadron Plus Strip (HPS) algorithm \cite{TAU-11-001}. 
The algorithm uses PF jets, reconstructed using anti-$k_{T}$ algorithm with a distance parameter R=0.5, as an initial seed. The algorithm first reconstructs the $\pi^{0}$ components of the $\tau_{h}$, then combines them with the charged hadrons to reconstruct the tau decay mode and calculate tau four-momentum and isolation quantities. Special attention is given to photon conversions in the CMS tracker material. The photons are reconstructed in ``strips", objects that are built out of electromagnetic particles (PF photons and electrons), to take into account the broadening of calorimeter energy deposit due to the bending of electron/positron tracks in the CMS magnetic field.  
Strips satisfying a minimum transverse momentum requirement of $p_{T}^{strip}~>$ 1 GeV/$c$ are finally combined with the charged hadrons to reconstruct individual decay modes. 
The following decay topologies are considered by the algorithm:
\begin{enumerate}
\item{{\it Single Hadron} corresponds to $h^{-}\nu_{\tau}$ and $h^{-}\pi^{0}\nu_{\tau}$ decays in which the neutral pions have too little energy to be reconstructed as strips.}
\item{{\it One hadron + one strip} reconstructs the decay mode $h^{-}\pi^{0}\nu_{\tau}$ in events where the photons from $\pi^{0}$ decay are close together on the calorimeter surface.}
\item{{\it One hadron + two strips} corresponds to the decay mode $h^{-}\pi^{0}\nu_{\tau}$ in events where photons from $\pi^{0}$ decay are well separated.} 
\item{{\it Three hadrons} corresponds to the decay mode $h^{-}h^{+}h^{-}\nu_{\tau}$. The three prong charged hadrons are required to come from the same secondary vertex.} 
\end{enumerate}
No separate decay topologies are considered for the $h^{-}\pi^{0}\pi^{0}$ and $h^{-}h^{+}h^{-}\pi^{0}\nu_{\tau}$ decay modes and they are reconstructed via the existing topologies. 
All charged hadrons and strips are required to be contained within a cone of size $\Delta R$~=~(2.8~GeV/$c$)/$p^{\tau_{h}}_{T}$, where $p^{\tau_{h}}_{T}$ is the transverse momentum of the reconstructed $\tau_{h}$.
The four-momenta of charged hadrons and strips are reconstructed according to the respective $\tau_{h}$ decay topology hypothesis, assuming all charged hadrons to be pions, and are required to be consistent with the masses of the intermediate meson resonances. 

The reconstructed $\tau_{h}$ candidates are required to be isolated.  
The isolation for $\tau_{h}$ candidates are computed using two approaches: cut based and multivariate analysis techniques.  
The cut based isolation criteria requires that the sum $p_{T}$ of charged hadrons and photons present within the isolation cone of size $\Delta R~=0.5$ around the direction of the $\tau_{h}$, apart from the $\tau_{h}$ decay products, is less than a certain threshold.  
In order to reduce the dependency of the isolation on the event pile-up, the charged hadrons which are originating from the same primary vertex as that of the leading charged hadron of the $\tau_{h}$ are considered for isolation.  
A correction ($\delta\beta$ correction) is applied to account for the neutral component of the pile-up in the isolation cone.  
The neutral component of the pile-up in the isolation cone is estimated as the sum $p_{T}$ of charged hadrons in the isolation cone originating from the pile-up vertices multiplied by the expected ratio of charged hadrons to the neutral hadrons.  
The estimated neutral component due to pile-up is subtracted from the isolation $\sum{p_{T}}$.  
By adjusting the threshold on isolation $\sum{p_{T}}$, three working points, ``loose'', ``medium'', and ``tight'' are defined.  
The ``loose'' working point corresponds to a probability of approximately 1\% for QCD jets to be misidentified as $\tau_{h}$.  
In the MVA approach, the isolation $\sum{p_{T}}$ is computed in annular rings around the $\tau_{h}$ candidates using the same PF particles used for cut based isolation. A boosted decision tree (BDT) is trained against the QCD jets. 
Three working points, ``loose'', ``medium'', and ``tight'' are defined depending on the cuts on the BDT output.  
 
\subsection{Performance of $\tau_{h}$ Identification} 
\begin{figure}[htp] 
%\begin{center} 
\centering 
\begin{tabular}{c} 
\includegraphics[width=0.45\textwidth]{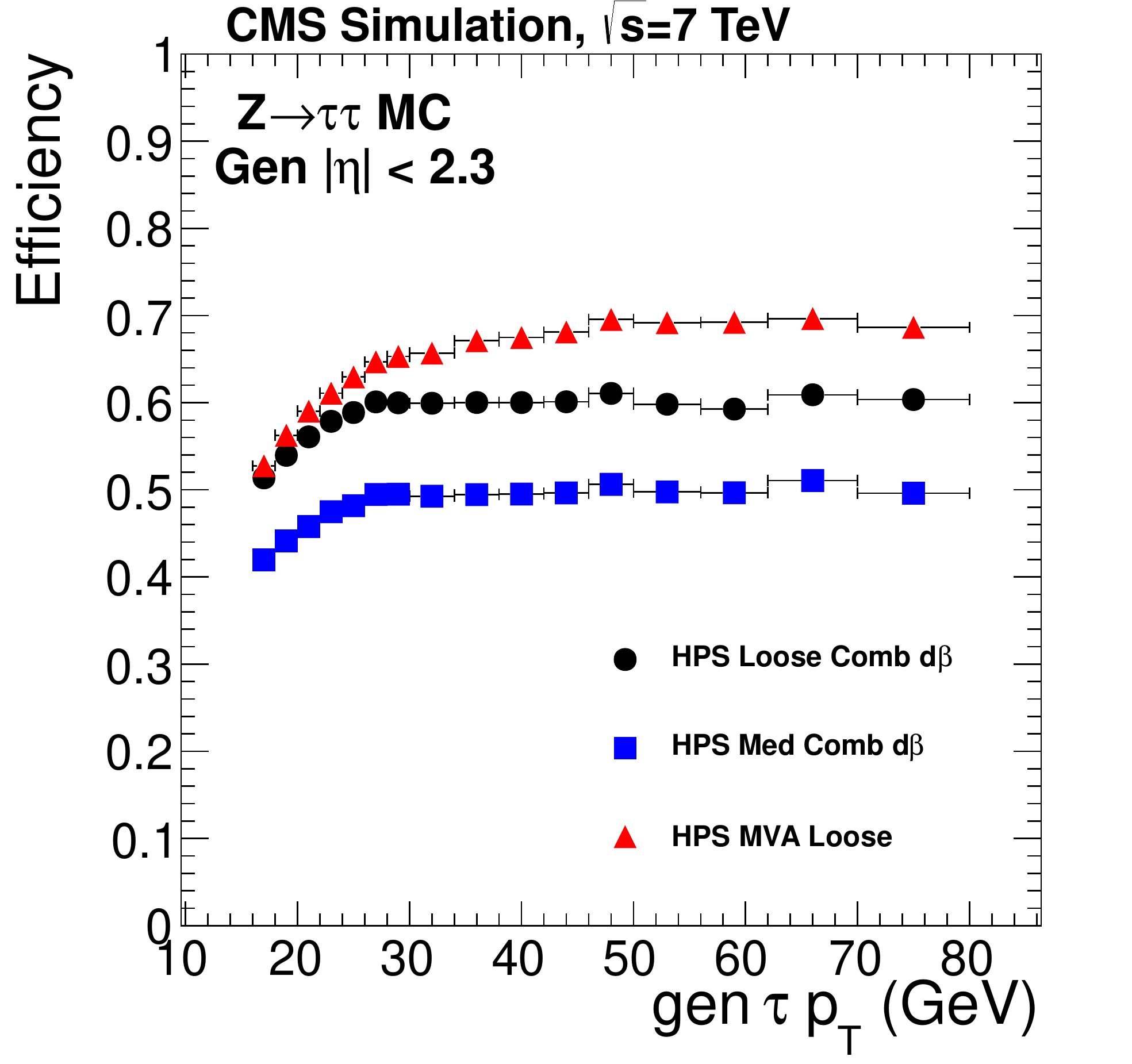} \hfill 
\includegraphics[width=0.45\textwidth,height=67mm]{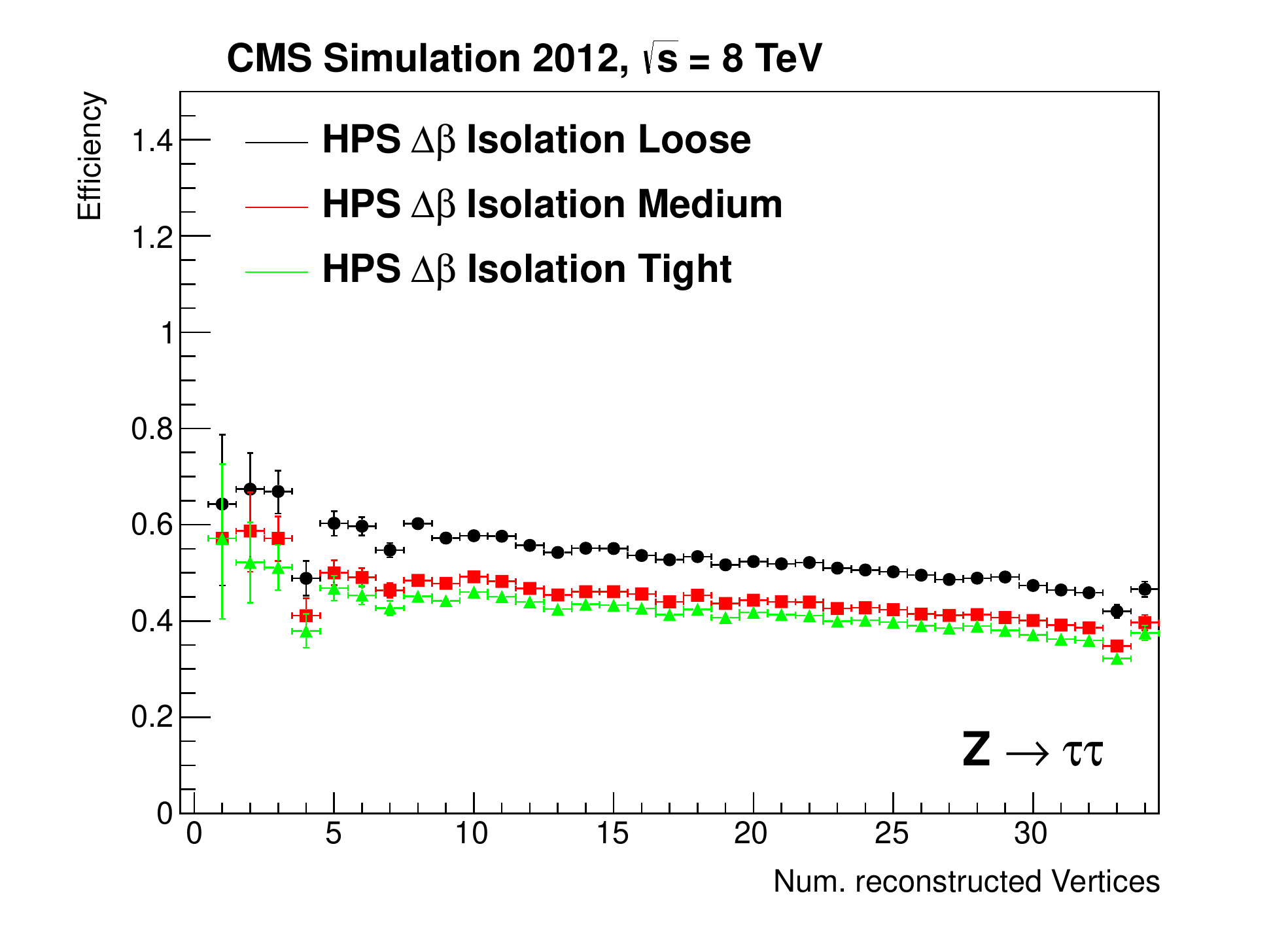}  
\end{tabular} 
\caption{Expected tau identification efficiency for different isolation working points as function of generated tau $p_{T}$ (left) and number of primary vertices (right). 
} 
\label{fig:TAU} 
%\end{center} 
\end{figure}

The efficiency of the $\tau_{h}$ reconstruction and identification is measured from data using a tag and probe method with a sample of Z$\rightarrow~\tau\tau~\rightarrow~\mu\tau_{h}$ events.  
The events are preselected using kinematic cuts and a set of requirements to suppress the background from Z$\rightarrow~\mu\mu$, W, and QCD events, but without applying the $\tau_{h}$-identification algorithm.  
The isolated muon is used as tag while an isolated jet candidate with a leading track above a certain threshold is considered as probe. The muon and the leading track in the jet are required to be of opposite charges.  
The HPS $\tau_{h}$-identification algorithm is applied to the jet in the preselected events.  
The invariant mass distribution of the muon-jet system for those events that pass or fail the $\tau_{h}$-identification are then fitted using signal and background distributions provided by MC simulation to extract the $\tau_{h}$ identification efficiency.   
The systematic uncertainties of the measured tau identification efficiency depend on the uncertainty arising from the template fit and the preselection efficiency. The total uncertainty is about 6-7\%. 
Fig.~\ref{fig:TAU} shows the expected $\tau_{h}$ identification efficiency as function of the generated visible $\tau$ $p_{T}$ estimated from $Z\rightarrow\tau\tau$ MC events for reconstructed $\tau_{h}$ threshold of 20 GeV/$c$. The efficiency has little dependence on the tau $p_{T}$ and the number of primary vertices, which shows the stability of the tau identification algorithm with event pile-up.  

The mis-identification probability of quark and gluon jets as $\tau_{h}$ have been measured from data using ``W+jets" and di-jet enriched events. 
%The jet $\rightarrow \tau_{h}$ mis-identification probability is found to be higher in ``W+jets" events compared to di-jet events, as expected, because of higher fraction of quark-jets. 
The jet $\rightarrow \tau_{h}$ mis-identification probability estimated from data and MC agree within approximately 20\%. 

\subsection{Discrimination Against Electrons and Muons}
Isolated electrons and muons passing the $\tau_{h}$ identification and isolation criteria are also important source of backgrounds in many analyses with $\tau_{h}$ in the final state. 
Muons reconstructed as $\tau_{h}$ are rejected using tracking information in the muon detectors and energy deposit in the calorimeter. 
With simple cut based criteria, the $\mu\rightarrow\tau_{h}$ mis-identification probability is reduced to less than 1\% while keeping $\sim$99\% efficiency for real taus. 
To reduce the e$\rightarrow\tau_{h}$ mis-identification probability a multivariate discriminant is used. A BDT is trained using input variables such as the $\tau$ position and momentum variables, cluster shape variables of PF photons, and tracking and cluster shape variables representing the compatibility of the leading charged hadron to be an electron. 
The e$\rightarrow\tau_{h}$ mis-identification probability is reduced to only a few percent while keeping the efficiency for real taus above 80\%.    
The e$\rightarrow\tau_{h}$ and $\mu\rightarrow\tau_{h}$ mis-identification probabilities are measured also from data using Z $\rightarrow$ ee and Z $\rightarrow~\mu\mu$ enriched events. 
 
\section{Reconstruction of Missing Transverse Momentum} %1+1/2 page 
The missing transverse momentum, $E_{T}^{miss}$, is reconstructed as the negative of the vector sum of the transverse momenta of all final-state particles reconstructed in the detector. 
There are three distinct algorithms developed in CMS to reconstruct $E_{T}^{miss}$: PF $E_{T}^{miss}$, Calo $E_{T}^{miss}$, and TC $E_{T}^{miss}$ \cite{JME-10-009}. PF $E_{T}^{miss}$ is calculated from the reconstructed PF particles, Calo $E_{T}^{miss}$ is calculated using the energies contained in calorimeter towers and their direction, relative to the center of the detector, to define pseudo-particles, and TC $E_{T}^{miss}$ is based on Calo $E_{T}^{miss}$, but the response and resolution is improved using tracks reconstructed in the inner tracker. 

A three-step correction is devised to remove the bias in the $E_{T}^{miss}$ scale due to the non-linearity of the response of the calorimeter for neutral and charged hadrons, caused by event pile-up, large bending of low $p_{T}$ tracks due to strong magnetic field in CMS, etc..
The correction procedure relies on the fact that $E_{T}^{miss}$ can be factorized into contributions from jets, isolated high $p_{T}$ photons, electrons, muons and unclustered energies.   
%The $E_{T}^{miss}$ is corrected for the effects due to the jet energy scale corrections using the so called "Type-I" correction:
The jet energy scale corrections are propagated to $E_{T}^{miss}$ using the so called "type-I" correction:

\begin{eqnarray}
E_{x,y}^{miss, corrected} =  E_{x,y}^{miss, raw} - \sum_{i}\left(p_{i, (x,y)}^{corrected~jets} - p_{i, (x,y)}^{raw~jets}\right)
\end{eqnarray}

In order to correct for the soft jets below the threshold used for "type-I" correction and energy deposits not clustered in any jet, a second correction can be applied to the unclustered energy, which is called "type-II" correction. This correction is obtained from Z$\rightarrow$ee  events, as discussed in \cite{JME-10-009}.
To reduce the dependency of $E_{T}^{miss}$ on event pile-up, a so called "type-0" correction has been developed only for PF $E_{T}^{miss}$. For each pile-up vertex the expected missing neutral momentum is calculated using an improved PF candidate to vertex association technique and added it vectorially to PF $E_{T}^{miss}$.   
 
\subsection{$E_{T}^{miss}$ Scale and Resolution}
\begin{figure}[htp]  
%\begin{center}  
\centering  
\begin{tabular}{c}  
\includegraphics[width=0.45\textwidth]{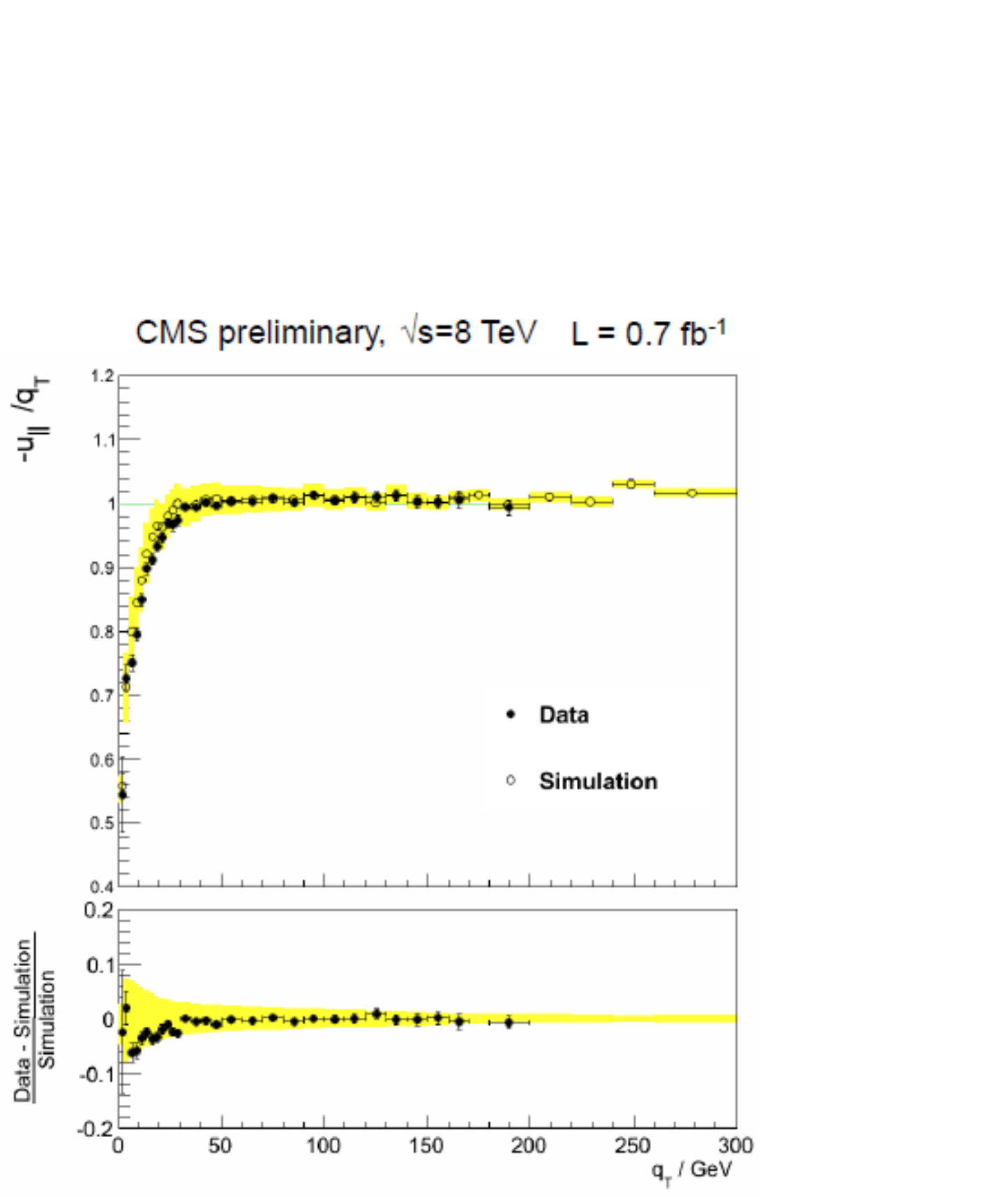} \hfill  
\includegraphics[width=0.45\textwidth]{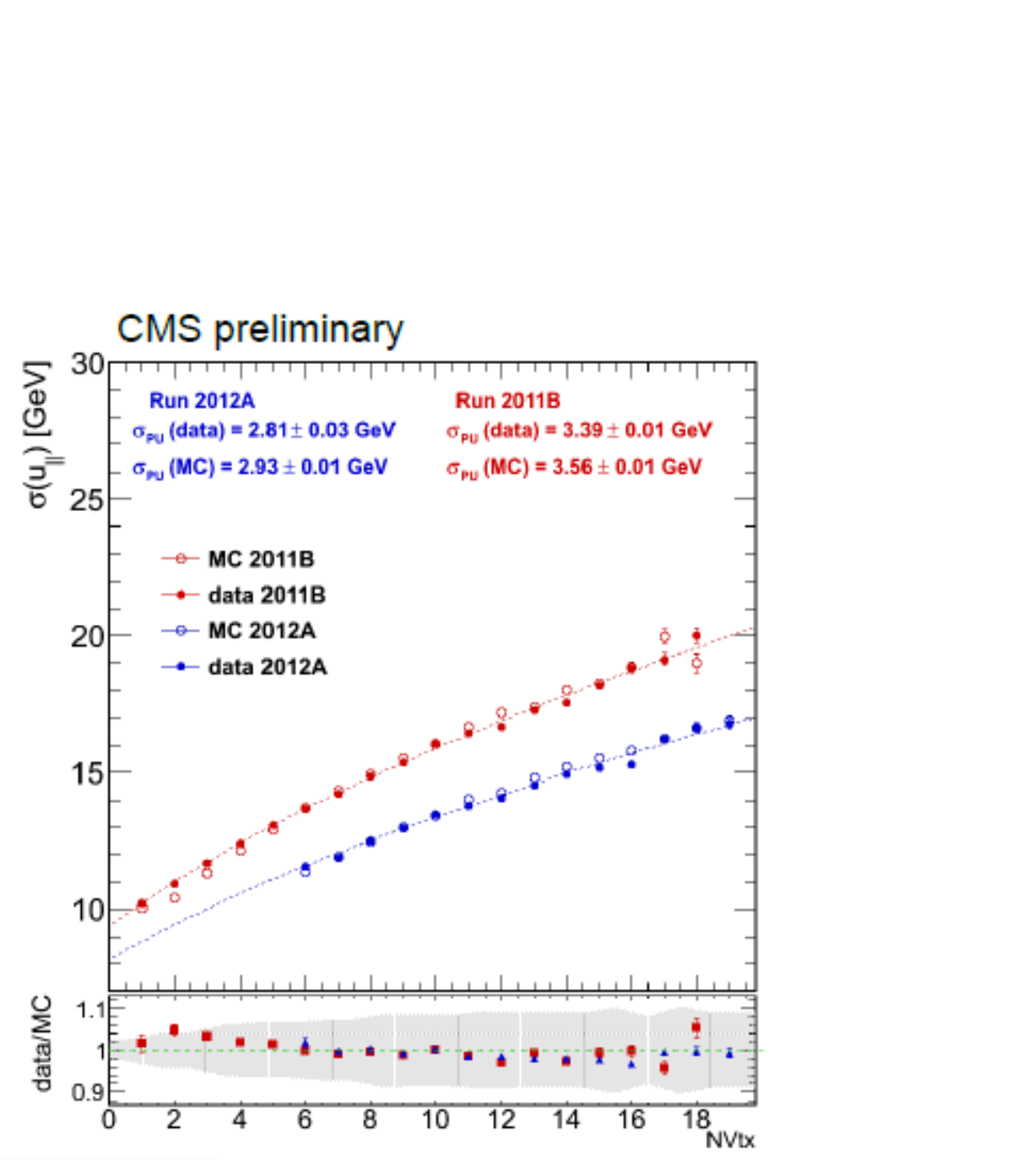}   
\end{tabular}  
\caption{Left: The $E_{T}^{miss}$ response measured in $Z~\rightarrow~\mu\mu$ events. 
Right: The $E_{T}^{miss}$ resolution as function of the number of primary vertices. The resolution is better in 2012 early run (2012A) data compared to 2011 data (2011B) due to improved energy reconstruction in electromagnetic and hadronic calorimeter to reduce the effects of the out-of-time pile-up interactions.   
}  
\label{fig:MET}  
%\end{center}  
\end{figure}

The performance of $E_{T}^{miss}$ is studied using events containing a Z boson where the $E_{T}^{miss}$ can be induced by removing the vector boson from the event. 
The well measured Z boson, from two electrons or two muons, provides the momentum scale, $q_{T}$ and a unique  event axis, $\hat{q}_{T}$.
The hadronic recoil, $\vec{u}_{T}$, is defined as the vector sum of the transverse momenta of all particles except the vector boson. 
Momentum conservation in the transverse plane requires $\vec{q}_{T}+\vec{u}_{T}=0$. 
The projection of the hadronic recoil onto the axis $\hat{q}_{T}$ yields two signed components, parallel ($u_{||}$) and perpendicular ($u_{\perp}$) to the event axis, where $u_{||}$ is typically negative.
The mean value of the scalar quantity $<u_{||}>/q_{T}$ is the correction factor required for $E_{T}^{miss}$ measurements and is closely related to jet energy scale corrections. $<u_{||}>/q_{T}$ is referred as response. 
Fig.~\ref{fig:MET} shows the response curve versus $q_{T}$ for 2012 early run data and MC simulation. Deviations of the response curve from unity probe the $E_{T}^{miss}$ response as function of $q_{T}$. 
The $E_{T}^{miss}$ resolution is measured as the RMS spread of $u_{||}$ and $u_{\perp}$ about their mean values, after correcting for the response. Fig.~\ref{fig:MET} shows the $E_{T}^{miss}$ resolution as function of the number of primary vertices. 
 
\section{b-Jet Identification} %1+1/2 page
The properties of the b-hadrons are used to identify hadronic jets originating from the fragmentation of b-quarks. 
These hadrons have relatively large masses, long lifetimes, and daughter particles with hard momentum spectra. 
Their semileptonic decays can be exploited as well. Various b-tagging algorithms used at CMS and their performance are briefly discussed here. 

\subsection{b-Tagging Algorithms}
CMS has developed a variety of algorithms to identify b-quarks based on variables such as the impact parameter of charged particle tracks, the properties of reconstructed decay vertices, and the presence of a lepton, or the combination of the above information \cite{BTV-12-001}.   
Each of these algorithms produces a single discriminator value for each jet. 
The minimum thresholds on these discriminators define loose (``L"), medium (``M"), and  tight (``T") working points corresponding to the mis-identification probability for light parton jets of approximately 10\%, 1\%, and 0.1\%, respectively, at an average jet $p_{T}$ of 80 GeV/$c$. 

The impact parameter (IP) of a track with respect to the primary vertex is calculated in three dimensions by taking the advantage of the excellent resolution of the pixel detector along the $z$ axis. 
The sign of the IP is defined as the sign of the scalar product of the vector pointing from the primary vertex to the point of closest approach with the jet direction. 
While the IP values of the tracks originating from the decay of particles traveling along the jet axis tend to have positive values, the IP of prompt tracks can have positive or negative values. 
The impact parameter significance $S_{IP}$, defined as the ratio of the IP to its estimated uncertainty, is used as a discriminating observable.  
The simplest algorithm based on the track impact parameter is called {\it Track Counting} (TC) algorithm which sorts tracks in a jet by decreasing values of IP significance. 
The {\it Track Counting High Efficiency} (TCHE) and {\it Track Counting High Purity} (TCHP) algorithms use the $S_{IP}$ of second and third ranked track as the discriminator value.    
The IP information of several tracks in a jet are also combined to provide better discriminating power.  
The {\it Jet Probability (JP)} algorithm uses an estimate of the likelihood that all tracks associated to the jet come from the primary vertex. The {\it Jet B Probability (JBP)} algorithm gives more weight to the tracks with the highest IP significance, up to a maximum of four such tracks, which matches the average number of reconstructed charged particles from the b-hadron decays.  
 
The presence of a secondary vertex provides the most powerful discrimination between b and non-b jets. 
The kinematic variables of the secondary vertex such as flight distance, direction, track multiplicity, mass or the energy are used in the b-tagging algorithms. 
The {\it Simple Secondary Vertex} (SSV) algorithm uses the significance of the flight distance, the ratio of flight distance to its estimated uncertainty, as the discriminating variable. 
A more complex algorithm, the {\it Combined Secondary Vertex} (CSV) algorithm, involves the use of secondary vertices together with track based lifetime information to provide the most efficient discrimination between b and non-b jets.   

\begin{figure}[htp]   
%\begin{center}   
\centering   
\begin{tabular}{c}   
\includegraphics[width=0.45\textwidth]{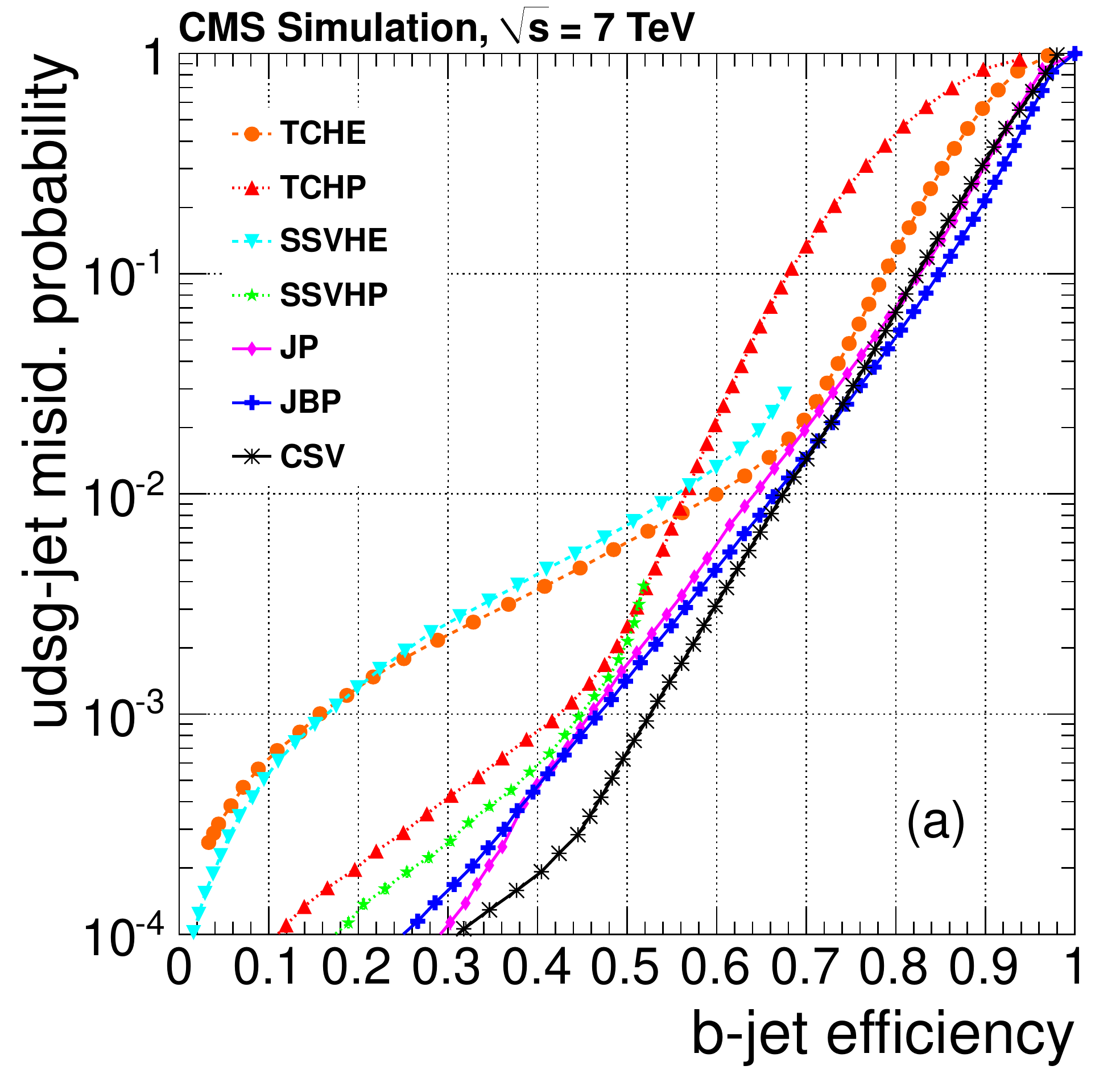} \hfill   
\includegraphics[width=0.45\textwidth]{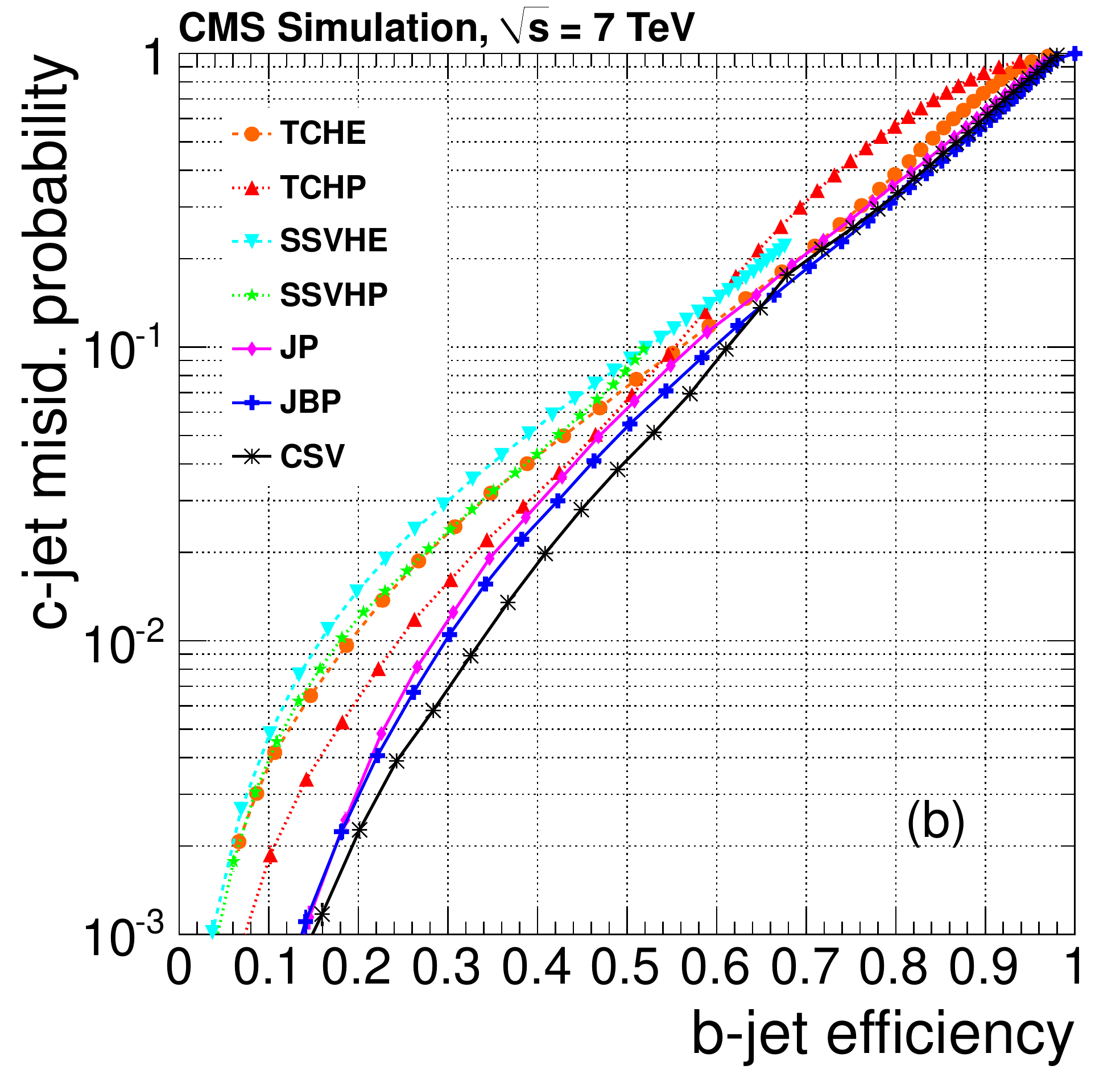}    
\end{tabular}   
\caption{Performance curves obtained from simulation for the algorithms described in the text.
(a) light-parton and (b) c-jet misidentification probabilities as a function of the b-jet efficiency}
\label{fig:BTag}   
%\end{center}   
\end{figure}   

\subsection{Performance}
The performance of the algorithms discussed above are summarized in Fig.~\ref{fig:BTag} where the misidentification probabilities predicted by simulation are plotted as function of b-jet efficiencies. 
As seen in Fig.~\ref{fig:BTag}, in the region where the misidentification probability is $\sim$10\% the JBP algorithm has high b-jet efficiency while in the region where misidentification probability is $<$1\% the CSV algorithm is the most efficient. 
Also the CSV algorithm provides the best c-jet rejection values in the high-purity region. 

The efficiency of the b-tagging algorithms are measured from data in order to reduce the dependency on simulation. 
There are a number of techniques which are applied to CMS data to measure the efficiencies using either dijet events or $t\bar{t}$ events~\cite{BTV-12-001}. 
The dijet events with a jet containing a muon within the jet cone (a ``muon jet") is used to measure the efficiency. 
A dijet sample with high b-jet purity is obtained by requiring that the ``away jet" (other than the muon jet) is b-tagged  using a lifetime based b-tagger. 
Various methods as discussed in \cite{BTV-12-001} are used to measure efficiency using muon-jet events. 
%The data to MC scale factors of b-tagging efficiency for all possible tagging algorithms used at CMS are estimated along with its systematic uncertainties as function of $p_{T}$ and $\eta$ of jets, which are then used in various physics analysis requiring b jets in the final state. 

The misidentification probability for the light-parton jets is also measured from data relying on the definition of inverted tagging algorithms, selecting non-b jets using the same variables and techniques as the standard versions. The negative tagger is computed from tracks  with negative impact parameter or from secondary vertices with negative decay length. 
The sample of negative tagged jets are enriched with light flavours. 
%The data to simulation scale factor for misidentification probability is measured along with its uncertainty which are then used in the analysis. 
  
\section{Reconstruction and Identification of Electrons and Muons} %1 page
Electrons are reconstructed by combining tracks in the inner tracker with the energy deposited in the electromagnetic calorimeter. Electron trajectories are reconstructed using a dedicated modeling of the electron energy loss due to bremsstrahlung radiation within the tracker material and are fitted with a Gaussian sum filter \cite{PAS-EGM-10-001}. 

The electrons are identified using track and cluster shape variables such as energy-momentum and the spatial match between the track and the ``supercluster"~\cite{PAS-EGM-10-001}, supercluster $\eta$ width, energy leaked to the hadronic calorimeter. Both cut based and MVA based algorithms are used to combine these variables for electron identification. 
A series of reference working points are defined depending on the efficiency of electron selection using Monte Carlo samples.
The electron isolation variables are computed in three sub-detectors: the tracker, the ECAL, and the HCAL. Transverse momentum/energy sums are evaluated in the region around the electron. 
To cope with the high event pile-up in 8 TeV data, isolations are computed using particle-flow candidates: charged hadrons, neutral hadrons and photons. Charged hadrons originating from the same primary vertex as the electron are considered for isolation. To account for the neutral energy due to pile-up in the isolation cone, a correction is applied using the pile-up energy density ($\rho$) estimated on an event by event basis using the jet area method as discussed in Sect.-\ref{ChapterJEC}.

Muons are reconstructed by combining information from muon chambers with that of the inner tracker \cite{MUO-10-004}. 
Tracks are first reconstructed independently in the inner-tracker (tracker track) and in the muon system (standalone muon track). 
The standalone muon tracks are matched to the tracker tracks and the hits are combined using a Kalman filter \cite{MUO-10-004} to reconstruct a ``global muon". 
Tracker only muons are also reconstructed by extrapolating tracks in the inner-tracker to the muon system. If at least one muon segment matches to the extrapolated track, the corresponding tracker track is considered as a ``tracker muon".
The muon identification is performed using several categories of muon identification algorithms based on variables such as track quality, compatibility of the calorimeter response with the muon hypothesis, and the presence of matched segments in the muon system. Particle-flow information is also used to improve the purity of the muon identification with substantial low fake rate from charged pions. Several baseline categories of muons are defined based on the analysis requirement. As discussed for electrons, the isolation for muons are also computed using particle-flow candidates with corrections to account for the neutral energy due to event pile-up.
    
\section{Summary}
The CMS experiment has developed excellent techniques for the reconstruction of physics objects by combining information from various components of the CMS detector. The use of the particle-flow technique has greatly improved the reconstruction of hadronic jets, missing transverse energy, and identification of hadronic decay of $\tau$ leptons. 
The particle-flow reconstruction also improved the robustness of the object identification methods against the high event pile-up. 
The reconstruction and identification of various objects and their performance in data have been briefly presented.

\end{document}